\documentclass[iop]{emulateapj}

\newcommand{\vw}{\ensuremath{v_\mathrm{w}}}
\newcommand{\Erad}{\ensuremath{E_\mathrm{rad}}}
\newcommand{\Em}{\ensuremath{E_\mathrm{m}}}
\newcommand{\Mm}{\ensuremath{M_\mathrm{m}}}
\newcommand{\Eex}{\ensuremath{E_\mathrm{ex}}}
\newcommand{\Mex}{\ensuremath{M_\mathrm{ex}}}
\newcommand{\Msh}{\ensuremath{M_\mathrm{sh}}}
\newcommand{\Mdots}{\ensuremath{\dot{M}_\mathrm{s}}}
\newcommand{\Mdotej}{\ensuremath{\dot{M}_\mathrm{ej}}}
\newcommand{\Msun}{\ensuremath{M_\odot}}
\newcommand{\Rs}{\ensuremath{R_\mathrm{sh}}}
\newcommand{\kmps}{\ensuremath{\mathrm{km~s^{-1}}}}
\newcommand{\Msunpyr}{\ensuremath{M_\odot~\mathrm{yr^{-1}}}}

\shorttitle{SN 2009\MakeLowercase{ip}}
\shortauthors{Moriya}

\begin{document}


\title{SN 2009\MakeLowercase{ip}: 
Constraining the latest explosion properties
by its late-phase light curve}

\author{Takashi J. Moriya}
\affil{Argelander Institute for Astronomy, University of Bonn,
    Auf dem H\"ugel 71, 53121 Bonn, Germany}
\email{moriyatk@astro.uni-bonn.de}
%
%



\begin{abstract}
We constrain the explosion and circumstellar properties at the 2012b
event of SN~2009ip based on its late-phase bolometric light curve
recently reported.
The explosion energy and ejected mass at the 2012b event are estimated
as 0.01~\Msun\ and $2\times 10^{49}$ erg, respectively.
The circumstellar medium is assumed to have two
 components: an inner shell and an outer wind.
The inner shell which is likely created at the 2012a event has 0.2~\Msun.
The outer wind is created by the wind mass loss before the 2012a mass
 ejection, and the progenitor is estimated to have had the mass-loss rate
about $0.1$~\Msunpyr\ with the wind velocity 550~\kmps\ before the 2012a event.
The estimated explosion energy and ejected mass indicate that
the 2012b event is not caused by a regular supernova.
\end{abstract}


\section{Introduction}
It is widely believed that the efficient conversion of kinetic
energy to radiation results in luminous transients.
Type~IIn supernovae (SNe),
which sometimes even become superluminous \citep[e.g.,][]{gal-yam2012},
are largely powered by the interaction between SN ejecta and circumstellar
media (CSM) \citep[e.g.,][]{moriya2014}.
There also exist bright transients called `SN impostors'
which are likely brightened by the collision of the material ejected
from the progenitor intermittently \citep[e.g.,][]{vandyk2000}.
SN impostors are not caused by the final SN
explosions of the progenitors, but they sometimes become as bright as SNe.
Type~IIn SNe and SN impostors are related to the unsolved
problems in stellar mass loss, and it is important to understand their
origins \citep[e.g.,][]{smith2014b,langer2012}.
They are also suggested to be an important high-energy
cosmic-ray producers \citep{murase2011,murase2014}.
In addition, they can play a role as a distance ladder \citep{potashov2013}
and can also be an important probe of the early Universe
\citep[e.g.,][]{cooke2009,tanaka2012}.

SN~2009ip is one of the most studied transients powered by the
interaction due to its activeness which kept us surprised
since 2009.
The progenitor of SN~2009ip got bright
and was assigned a SN name in August 2009 \citep{maza2009}.
However, the subsequent observations revealed that it was not a genuine
SN and the progenitor remained at the location \citep{smith2010,foley2011}.
The progenitor mass is estimated to be above $\sim 60$~\Msun, and
the observed brightening is considered to be a SN impostor from
a luminous blue variable star \citep{smith2010,foley2011}.
The progenitor experienced several rebrightening since 2009 \citep{pastorello2013}.

SN~2009ip showed drastic changes in 2012 (Fig.~\ref{lc}).
In August 2012, SN~2009ip started to be bright again and faded
temporarily after about 40 days (the 2012a event, see Fig.~\ref{lc}).
Then, it became bright again and reached the peak luminosity of
$8\times 10^{42}~\mathrm{erg~s^{-1}}$ \citep{pastorello2013},
which is comparable to those observed in SNe (the 2012b event, Fig.~\ref{lc}).
The observations during and after the 2012 events of SN~2009ip are reported by
many authors
\citep{prieto2013,pastorello2013,fraser2013,fraser2015,margutti2014,
ofek2013,smith2013,smith2014,mauerhan2013,mauerhan2014,levesque2014,graham2014,
martin2015,fox2015}.

The origin of the final luminosity increase observed so far (the 2012b event)
has been largely debated.
The peak luminosity which is comparable to those of SNe and the broad
spectral lines led to the suggestion that the 2012b event is triggered
by the final SN explosion of the progenitor
\citep[e.g.,][]{mauerhan2013,smith2014,baklanov2013,ouyed2013}.
The 2012a event, which is similar to the precursor of SN~2010mc
\citep{ofek2013b},
is linked to the pre-SN mass ejection probably caused by the violent late-phase
nuclear burning (e.g., \citealt{quataert2012}, see also \citealt{moriya2014c}).
On the other hand, it is also suggested that a SN event may not
be required to explain the large luminosity observed in the 2012b event
because of the efficient conversion from the kinetic energy to radiation.
Thus, the 2012b event may not be caused by the final SN explosion
of the progenitor
\citep[e.g.,][]{pastorello2013,fraser2013,fraser2015,margutti2014}.
For example, \citet{soker2013,kashi2013,tsebrenko2013} related the 2012 events to
the merger of massive stars.

The aim of this Letter is to reveal the unresolved origin of the
enigmatic 2012b event by modeling its late-phase bolometric light curve (LC)
until about 750 days after the 2012b event recently reported
by \citet{fraser2015}.
We constrain the explosion and circumstellar properties at the 2012b event
by using an analytic bolometric LC model developed by \citet{moriya2013}.
We begin this Letter by briefly summarizing our LC model in the next section.

\section{Light-curve model}\label{lcmodel}
\subsection{Assumptions}
\begin{figure}
\epsscale{1.0}
\plotone{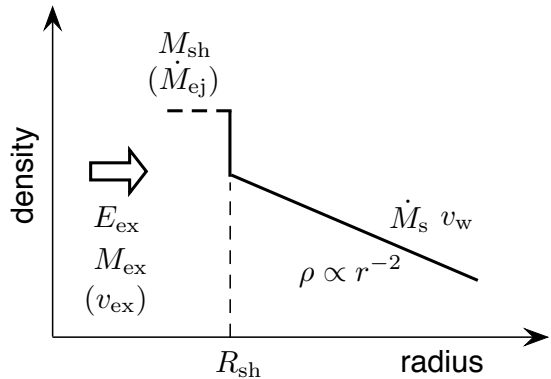}
\caption{
Schematic illustration of the explosion and circumstellar properties
assumed in our model. The ejecta with the kinetic energy \Eex\ and the mass
 \Mex\ exploded at the 2012b event first collides with the inner shell with
 \Msh\ created at the 2012a event with the mass-loss rate \Mdotej.
Then, the thin shock containing both \Mex\ and \Msh\ interacts with
the outer wind created by the progenitor prior to the 2012a event
with the mass-loss rate \Mdots\ and velocity \vw.
}
\label{picture}
\end{figure}

We assume a progenitor system schematically shown in
Fig.~\ref{picture} led to the 2012b and later event of SN~2009ip.
The event is assumed to be caused by an ejection of the mass \Mex\
with the kinetic energy \Eex\ in a dense CSM.
The dense CSM are assumed to have two components:
an inner shell and an outer regular wind.
The inner shell is presumed to be created by the mass ejection
during the 2012a event with the mass-loss rate \Mdotej,
and it is assumed to have the mass \Msh.
We note that this kind of shells may also be formed by
the confinement of the stellar wind \citep{mackey2014}.
The outer wind structure is assumed to be created with wind mass loss
with the mass-loss rate \Mdots\ and velocity \vw.
Although the observed luminosity fluctuations before the 2012a event
indicate that the CSM may not be smooth,
we assume that the overall density structure is approximately
proportional to $r^{-2}$.
Based on the observations of SN~2009ip in 2009, we assume that
$\vw=550~\kmps$ \citep{smith2010,foley2011}.
The inner and outer components are separated at \Rs.

We assume that the inner shell is responsible for the rise and
decline observed until $\sim 100$ days since the 2012b event
(`early phase', Section~\ref{earlyphase}).
Then, a shock containing both \Mex\ and \Msh\ is assumed to propagate
in the outer wind.
The shock is assumed to be thin due to the efficient radiative cooling.
The continuous interaction between the shock and the outer wind is
assumed to be responsible for 
the later ($\gtrsim 200$ days) LC reported by \citet{fraser2015}
(`late phase', Section~\ref{latephase}).

Finally, we note that the spherical symmetry is assumed in our analytic
model. Some observations of the 2012b event are suggested to indicate that
the CSM is aspherical \citep[e.g.,][]{levesque2014,mauerhan2014,graham2014}.
However, the mass-loss rates obtained by assuming the spherical symmetry
are still likely to be good estimates for the actual mass-loss rates
in the aspherical systems \citep{moriya2014}.
In addition, the conversion efficiency discussed below also partly
contains the effect of the asphericity \citep{moriya2014}.

\begin{figure}
\epsscale{1.2}
\plotone{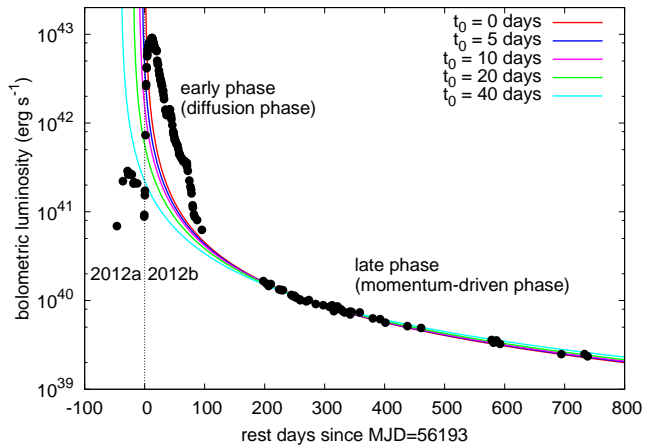}
\caption{
Bolometric LC of SN~2009ip in 2012 and later shown in \citet{fraser2015}.
The origin of time is set at the beginning of the 2012b event.
We also show the results of the LC fitting with the
function $L=L_1(t+t_0)^{-1.5}$ after 150 days since the beginning of the
 2012b event.
}
\label{lc}
\end{figure}

\subsection{Early phase (diffusion phase)}\label{earlyphase}
We assume that the early phase is caused by the diffusion in the inner
shell (see also \citealt{margutti2014,ofek2013}).
If we assume that the shell has an average density $\overline{\rho}$,
the diffusion time $t_\mathrm{diff}$ in the shell,
which corresponds to the rise time of the LC, is expressed as
\begin{equation}
 t_\mathrm{diff}\simeq\frac{\kappa \overline{\rho} \Rs^2}{c},\label{eqdiff}
\end{equation}
where $\kappa$ is opacity and $c$ is the speed of light.
We assume $\kappa=0.34$ $\mathrm{cm^2~g^{-1}}$ in this study.
Using the diffusion time, \Msh\ can be estimated as
\begin{equation}
 \Msh \simeq \frac{4}{3}\pi\overline{\rho}\Rs^3
=\frac{4\pi c t_\mathrm{diff}\Rs}{3\kappa}.
\end{equation}

As the rise time and the $e$-folding time of the later luminosity
decline are both about 14 days \citep[e.g.,][]{pastorello2013,moriya2014},
a shock breakout is likely occurred in the shell \citep{margutti2014,ofek2013}.
This indicates that the entire shell is shocked at the LC peak.
Thus, we presume that the blackbody radius at the LC peak of the 2012b
event corresponds to \Rs\ and we set $\Rs\simeq10^{15}$~cm \citep{margutti2014}.
Assuming $t_\mathrm{diff}\simeq14~\mathrm{days}$ and $\Rs\simeq10^{15}$~cm, we
obtain $\Msh\simeq 0.22~\Msun$.
The estimated mass is consistent with those obtained in the previous
studies ($\sim 0.1~\Msun$, e.g., \citealt{fraser2013,margutti2014}).
If the shell is ejected during the 2012a event which lasted for
$\sim 40$ days, the mass-loss rate during the 2012a event becomes
$\Mdotej\simeq 2~\Msunpyr$.

The explosion properties (\Eex\ and \Mex) can be
related to the total radiation energy emitted during the 2012b event.
The conservation of momentum and energy results in \citep[e.g.,][]{moriya2013b}
\begin{equation}
\Erad = \frac{\Msh}{\Mex+\Msh}\Eex.
\label{energye}
\end{equation}
The total energy during the 2012b event is estimated to be 
$\Erad\simeq 2\times 10^{49}$ erg \citep{fraser2013,margutti2014}.

\subsection{Late phase (momentum-driven phase)}\label{latephase}
\citet{fraser2015} recently reported the late phase LCs until about
750 days after the 2012b event and constructed a bolometric LC (Fig.~\ref{lc}).
The late bolometric LC is shown to evolve with a power law
$L=L_1 t^{-\alpha}$, where $t$ is time since the explosion.
\citet{fraser2015} fitted the power-law function by setting
$t=0$ at the beginning of the 2012a event and obtained
$\alpha=1.74$ as the best fit parameter.
However, we here assume that the inner explosion causing the 2012b event
occurred after the 2012a event. If we set $t=0$ at $\mathrm{MJD}=56193$ when
the 2012b event began, we obtain $\alpha=1.44$ as the best fit parameter.

What is surprising is that $\alpha$ is near 1.5 and it is significantly
larger than 1.0. If the interaction between SN ejecta
and the dense wind is still ongoing,
$\alpha$ is expected to be significantly
below 1.0 when the wind is almost steady 
because of the continuous momentum injection from the SN ejecta
\citep[e.g.,][]{moriya2013,ofek2014}.
For example, in the case of Type~IIn SN~2010jl, the bolometric LC
follows a power law with $\alpha<1$ even at around 600 days after the
explosion \citep[e.g.,][]{maeda2013}, although there exist some
arguments for the LC interpretation \citep{fransson2014}.

The large $\alpha$ near 1.5 in the LC after about 200 days
since the 2012b explosion indicates that
the shock proceeding in the wind is already
in the \textit{momentum-driven} phase at about 200 days.
The momentum-driven phase (also called `snow-plow phase')
is the phase when the momentum injection to the shock has already terminated
and the shock moves only with the momentum injected to the shock previously
\citep[see, e.g., ][]{svirski2012,ofek2014,moriya2014b}.
\citet{moriya2013} obtained the luminosity evolution during
the momentum-driven phase in the steady wind as
\begin{equation}
 L=\frac{\epsilon}{2}\frac{\Mdots}{\vw}
\left(\frac{2\Em}{\Mm}\right)^{\frac{3}{2}}
\left[
1+2\frac{\Mdots}{\vw}
\left(\frac{2\Em}{\Mm^3}\right)^{\frac{1}{2}}t
\right]^{-\frac{3}{2}},
\label{Lmomentum}
\end{equation}
where
$\epsilon$ is the conversion efficiency from the kinetic energy to
radiation, and
\Em\ and \Mm\ is the energy and mass released inside the wind, respectively.
In the system we are interested in (Fig.~\ref{picture}),
the shock propagates in the outer wind component after it has passed through
the inner shell. Thus, we can set
\begin{equation}
\Mm = \Mex+\Msh,\label{masse}
\end{equation}
from the conservation of mass, and
\begin{equation}
\Mm\Em = \Mex\Eex,\label{momentume}
\end{equation}
from the conservation of momentum.

The luminosity evolution in the momentum-driven phase
(Eq.~\ref{Lmomentum}) can be separated into two parts.
At first, when $2(\Mdots/\vw)(2\Em/\Mm^3)^{1/2}t \lesssim 1$,
the luminosity is constant. This is because the wind mass
swept by the shock is much smaller than the initial injected mass (\Mm)
and the shock freely expands with a constant velocity.
Then, when 
\begin{equation}
 2\frac{\Mdots}{\vw}
\left(\frac{2\Em}{\Mm^3}\right)^{\frac{1}{2}}t \gtrsim 1,
\label{latecondition}
\end{equation}
starts to hold after the shock has swept a large amount of the wind
(see \citealt{moriya2014b} for detailed discussion on this condition),
the luminosity evolves as $L=L_1 t^{-1.5}$, where
\begin{equation}
 L_1=2^{-\frac{7}{4}} \epsilon \left(\frac{\Mdots}{\vw}\right)^{-\frac{1}{2}}
\Em^{\frac{3}{4}}\Mm^{\frac{3}{4}}.
\label{lume}
\end{equation}
Using Eq.~(\ref{lume}), the condition~(\ref{latecondition}) can be used
to constrain \Mm, i.e.,
\begin{equation}
\Mm\lesssim 2^{\frac{4}{3}}\epsilon^{-\frac{1}{3}}L_1^{\frac{1}{3}}
\left(\frac{\Mdots}{\vw}\right)^{\frac{2}{3}}t^{\frac{1}{2}}.
\label{lesssimMm}
\end{equation}
The late-phase bolometric LC of SN~2009ip indicates that
the condition (\ref{lesssimMm})
is satisfied at least about 200 days after the explosion.

In Fig.~\ref{lc}, we show the results of fitting of the function
$L=L_1(t+t_0)^{-1.5}$ to the bolometric LC after 150 days
since the beginning of the 2012b event
with several different explosion times ($t_0$) relative to the beginning
of the 2012b event ($\mathrm{MJD}=56193$).
Although $\alpha=1.5$ is fixed in the fitting,
decent fits to the bolometric LC are obtained.
The best $L_1$ is found as
$1.15\times 10^{51}$ ($t_0=0$~days),
$1.18\times 10^{51}$ ($t_0=5$~days),
$1.22\times 10^{51}$ ($t_0=10$~days), 
$1.29\times 10^{51}$ ($t_0=20$~days), and
$1.42\times 10^{51}$ ($t_0=40$~days) in the cgs unit.
Since $L_1$ does not depend strongly on $t_0$, we use 
$L_1=1.15\times 10^{51}$~cgs ($t_0=0$~days) as a representative value
in the following discussion.

Finally, the mass ejected at the inner explosion (\Mex) can be expressed as
\begin{equation}
 \Mex =
-\frac{\Msh}{2}+\frac{1}{2}\left[
\Msh^2+2^{\frac{13}{3}}\epsilon^{-\frac{4}{3}}L_1^{\frac{4}{3}}
\left(\frac{\Mdots}{\vw}\right)^{\frac{2}{3}}
\Msh\Erad^{-1}
\right]^{\frac{1}{2}},
\label{eqMex}
\end{equation}
using Eqs.~(\ref{energye}), (\ref{masse}), (\ref{momentume}), and (\ref{lume}).

\section{Explosion and circumstellar properties
at the 2012\MakeLowercase{b} event of SN~2009\MakeLowercase{ip}}\label{originof09ip}
We now look into the explosion and circumstellar properties of the final
explosive event observed in SN~2009ip so far.
We have already constrained the inner shell mass in the previous section
($\Msh\simeq 0.22$ \Msun).
The wind velocity is fixed to $\vw=550~\kmps$.
We also set $\epsilon =0.3$, which is typically found in
Type~IIn SN studies \citep[e.g.,][]{vanmarle2010,fransson2014}.
The conversion efficiency is related to the physical properties of
radiative shocks, and it is not likely to be altered by the origins of
the explosions inside. Thus, we use a typical value found in the SN
studies here.
The conversion efficiency can be reduced by, e.g., multi-dimensional motions
and asphericity \citep[e.g.,][]{moriya2013b}.
The observational information we have is
$\Erad\simeq 2\times 10^{49}$~erg and $L_1\simeq 1.15\times 10^{51}$~cgs.
We first assume several mass-loss rates for the outer wind
$(\Mdots=10^{-1}, 10^{-2}, 10^{-3}~\Msunpyr)$ and give constraints
on the other parameters for the assumed mass-loss rates.

The explosion properties at the 2012b event can be easily constrained
with the formulae derived in Section~\ref{lcmodel}.
First, the ejected mass \Mex\ can be constrained with Eq.~(\ref{eqMex}).
Then, the explosion energy \Eex\ can be estimated with Eq.~(\ref{energye})
assuming the obtained \Mex\ and \Msh.
Finally, we need to check if the condition (\ref{lesssimMm}) holds for the
estimated parameters for consistency.

Table~\ref{table} summarizes the estimated parameters for SN~2009ip.
We can first find that \Mex\ is much smaller than \Msh.
This means that most of the kinetic energy released at the 2012b event
is converted to radiation energy (Eq.~\ref{energye}).
In other words, Eq.~(\ref{energye}) indicates that
$\Erad\simeq \Eex$, and only small amount of the released kinetic energy is
available for the late phase. The remaining kinetic energy (\Em) is
only below 10\% of \Eex\ (Table~\ref{table}).
The total amount of energy radiated after 100 days
obtained by assuming $L=L_1t^{-1.5}$ is $7.8\times 10^{47}$~erg,
and only the $10^{-1}$~\Msunpyr\ model is consistent with the
total radiated energy.

The fact that the late-phase bolometric LC roughly follows $\propto t^{-1.5}$
gives the constraint~(\ref{lesssimMm}).
Assuming $t\simeq 200$~days in Eq.~(\ref{lesssimMm}),
we obtain the following constraint:
\begin{eqnarray}
\Mex+\Msh &\lesssim&
\left\{ \begin{array}{lll}
0.2\ \Msun & (\Mdots=10^{-1}~\Msunpyr),\\ 
0.04\ \Msun & (\Mdots=10^{-2}~\Msunpyr),\\ 
0.009\ \Msun & (\Mdots=10^{-3}~\Msunpyr).\\ 
\end{array} \right. \label{d}
\end{eqnarray}
Because \Msh\ is estimated to be 0.22 \Msun\ and $\Msh\gg\Mex$,
this constraint also indicates \Mdots\ is around $10^{-1}~\Msunpyr$.
The estimated mass-loss rate is consistent with those estimated
by the H$\alpha$ luminosity \citep{fraser2013,ofek2013}, while
it is lower than that estimated by \citet{ofek2013} based on
the multi-wavelength observations.
Combining above all,
we suggest $\Mex\simeq 0.011~\Msun$ and $\Eex\simeq 2.1\times 10^{49}$~erg
as the explosion properties at the 2012b event.

\begin{table}
\begin{center}
\caption{Estimated explosion and circumstellar properties \\
at the 2012b event}
\label{table}
\begin{tabular}{cccccc}
\tableline\tableline
\Mdots\tablenotemark{a} & \Eex & \Mex & \Msh & \Em  & $v_\mathrm{ex}$ \\
$\Msun~\mathrm{yr^{-1}}$ & $10^{49}$ erg & \Msun & \Msun  &
$10^{47}$ erg & $10^4~\mathrm{km~s^{-1}}$ \\
\tableline
$10^{-1}$ & 2.1   & $1.1\times 10^{-2}$ & 0.22 & 10   & 0.98 \\
$10^{-2}$ & 2.02  & $2.5\times 10^{-3}$ & 0.22 & 2.2  & 2.0 \\
$10^{-3}$ & 2.005 & $5.3\times 10^{-4}$ & 0.22 & 0.48 & 4.3 \\
\tableline
\end{tabular}
\tablenotetext{1}{$\vw=550~\kmps$}
\end{center}
\end{table}

The estimated \Eex\ and \Mex\ with $\Mdots=10^{-1}~\Msunpyr$
are also consistent with the interpretation that the shock breakout
occurred in the inner shell
at the 2012b event (Section~\ref{earlyphase}).
The explosion velocity $v_\mathrm{ex}\equiv\sqrt{2\Eex/\Mex}$ is shown
in Table~\ref{table}. The explosion velocity indicates that
the shock breakout occurs where the optical depth is
$\sim c/v_\mathrm{ex}\simeq 30$ \citep[e.g.,][]{weaver1976}.
If we use $\overline{\rho}\simeq 10^{-13}~\mathrm{g~cm^{-3}}$
estimated by Eq.~(\ref{eqdiff}), the shock breakout 
occurs at $\simeq 2\times 10^{14}$ cm. This radius is 
near $\simeq 4-5\times 10^{14}$~cm, which is the smallest photospheric radius
observed at the beginning of the 2012b event \citep{margutti2014}.

The estimated average ejecta velocity of $v_\mathrm{ex}\simeq 10^4~\kmps$
is also consistent with the broad spectra observed in the 2012b event
of SN~2009ip indicating the ejecta velocity of $\simeq 10^{4}$ \kmps\
\citep[e.g.,][]{fraser2013}.
\citet{pastorello2013} reported similarly broad spectral lines
in SN~2009ip before the 2012 events, and this kind of broad lines
are known to be associated with non-SN events.

The estimated explosion energy and mass at the 2012b event
are not those of regular SNe which typically have
$\sim 10^{51}$~erg and $\sim 1-10$ \Msun.
In particular, the progenitor of SN~2009ip is likely heavier than
60~\Msun \citep{smith2010,foley2011},
and the estimated small ejecta mass is inconsistent with its successful
SN explosion.
Even if the ejecta has $\sim 10^{51}$~erg and only the ejecta within
a certain direction interacted with the dense CSM to emit only 1\% of
the kinetic energy ($\sim 10^{49}$~erg, e.g., \citealt{smith2014}),
the estimated ejecta mass is much smaller than 1\% of the progenitor mass.
Thus, the explosion occurred inside at the 2012b event
is not likely related to a regular SN.
However, some SNe may have properties similar to
those estimated here, and 
we cannot conclude for sure if the core collapse of the progenitor
occurred or not.
For example, a SN with large fallback can have a small amount of
ejecta with a small kinetic energy \citep[e.g.,][]{moriya2010}.
Low energy and small mass ejection is also predicted
to be caused by failed SNe \citep[][]{nadezhin1980,lovegrove2013}.
Massive star mergers may also have a similar fast mass ejection
\citep[e.g.,][]{soker2013}.

\section{Conclusions}\label{conclusions}
We have estimated the explosion and circumstellar properties
at the 2012b event of SN~2009ip based on the late-phase bolometric LC
recently reported by \citet{fraser2015}.
The bolometric LC roughly follows $L\propto t^{-1.5}$ at least from
about 200 days after the 2012b event (Fig.~\ref{lc}),
and it indicates that the shock is already
at the momentum-driven phase without any momentum-injection
from inside at that time.
Thus, we use an analytic bolometric LC model
for the momentum-driven phase
to estimate the explosion and circumstellar properties.
We assume that an explosion occurred inside the circumstellar medium
with two components: an inner shell and an outer wind
(Fig.~\ref{picture}). The inner shell is supposed to be created during
the 2012a event of SN~2009ip, while the outer wind is made by the
wind mass loss of the progenitor prior to the 2012a event.

Combining all the bolometric LC information available after the 2012b event,
we suggest that
an explosion with the energy $2.1\times 10^{49}$~erg and the mass 0.011~\Msun\
occurred at the 2012b event of SN~2009ip.
The ejecta first collided to the inner shell whose mass is
estimated to be 0.22~\Msun.
The collision between the ejecta and the inner shell is responsible
for the early LC during the 2012b event which is dominated
by the photon diffusion after the shock breakout in the shell.
The thin shock made by the efficient cooling, which contains the mass of
both the ejecta and the inner shell, continues to travel in the outer wind,
powering the late-phase LC of SN~2009ip.
The total radiated energy and
the fact that the bolometric LC is already at the momentum-driven phase
with $L\propto t^{-1.5}$ at about 200 days after the explosion 
indicate that the mass-loss rate of the progenitor prior to the 2012a event
is about $0.1~\Msunpyr$ with the wind velocity 550~\kmps.

The estimated explosion properties are not those of regular SNe.
Thus, the explosion at the 2012b event is not related to a regular SN.
It is likely to be a non-SN explosive event or a peculiar SN like those
accompanied with large fallback or caused by failed SN explosions.

\acknowledgments

I would like to thank the referee for the constructive comments and
Morgan Fraser for sending me the electric data of the bolometric LC
of SN~2009ip.
The author is supported by Japan Society for the Promotion of
 Science Postdoctoral Fellowships for Research Abroad
 (26\textperiodcentered 51).

\end{document}